\documentclass[12pt,preprint]{aastex}

\newcommand{\aas}{AAS}

\begin{document}

\title{Photometric Selection of QSO Candidates From GALEX Sources}

\author{David W. Atlee} \and \author{Andrew Gould}
\email{atlee@astronomy.ohio-state.edu}
\affil{Department of Astronomy, The Ohio State University, Columbus, OH 43210}

\begin{abstract}
We present a catalog of $36\,120$ QSO candidates from the GALaxy Evolution 
EXplorer (GALEX) Release Two (GR2) UV catalog and the USNO-A2.0 
optical catalog.  The selection criteria are established using known quasars 
from the Sloan Digital Sky Survey (SDSS).  The SDSS sample is then used to 
assign individual probabilities to our GALEX-USNO candidates.  The mean
probability is $\sim50\%$, and would rise to $\sim65\%$ if better morphological
information than that from USNO were available to eliminate galaxies.  
The sample is $\sim40\%$ complete for $i\leq19.1$.  Candidates are
cross-identified in 2MASS, FIRST, SDSS, and XMM-Newton Slewing Survey (XMMSL1),
whenever such counterparts
exist.  The present catalog covers the 8000 deg$^{2}$ of GR2 lying above
$|b|=25^{\circ}$, but can be extended to all 24\,000 deg$^{2}$ that 
satisfy this criterion as new GALEX data become available.
\end{abstract}


\section{Introduction}\label{intro}
One of our most important observational tools is a well-determined system of 
coordinates against which to measure the position of any given target.
QSOs provide a natural means to establish an absolute frame
of reference.  The major drawback of using such a system today is
the strong variation in the density of known QSOs across the sky: the great
majority of known QSOs \citep{vero06} come from
the Sloan Digital Sky Survey (SDSS), which is spectroscopically
identifying quasars with $i\leq19.1$ with unprecedented completeness over the
$\sim$10\,000 deg$^{2}$ of the north Galactic cap \citep{rich02}.  The 
remaining $\sim$75$\%$ of the sky has a much lower density of known quasars.

There has also been recent interest in the spatial correlation function of
QSOs (see, for example, \citealt{yang06}).  While the SDSS sample covers a 
sufficiently large area on the sky to avoid biases due to cosmic variance,
its limited coverage of the south Galactic sky means that it could miss
potentially interesting features, while a catalog of QSOs constructed from
our candidate set would not suffer from this drawback.  Furthermore, a
magnitude selected subset of a QSO catalog from the complete Galaxy Evolution
Explorer (GALEX) All-sky Imaging Survey (AIS) could provide uniform coverage of
the entire sky away from the Galactic disk.

In this paper, we develop selection criteria to identify QSO 
candidates using UV and optical photometry of sources drawn from GALEX Release 
Two (GR2) and USNO-A2.0 \citep{mone98}, respectively.  Our use of
optical photometry from USNO means that our selection criteria can be applied
to sources in future GALEX releases, eventually extending our sky coverage
to the 24\,000 deg$^{2}$ above $|b|=25^{\circ}$, significantly larger than the
area covered by SDSS.
We tune our selection criteria by comparing to the colors of known quasars from
SDSS Data Release Four (DR4); we also evaluate our 
selection efficiency and completeness using DR4.  (For details concerning the
procedures used by SDSS to select and verify candidate QSOs, 
see \citealt{schn05}.)

\citet{bian05} have previously identified QSO candidates by matching GALEX and
SDSS sources over a small area of the sky (92 ${\rm deg}^{2}$).  
Because their fields overlap
SDSS with its superior photometry and morphological information, 
their efficiency and completeness are
much better than ours, but at the cost of restricting their 
sample to regions of the sky that already have the densest quasar coverage.
In a subsequent paper (\citealt{bian06}), the area considered has been expanded
to 353 ${\rm deg}^{2}$, examining sources from GALEX GR1 and SDSS DR3.

\section{Input Catalogs}\label{input}
GALEX records magnitudes in two
bandpasses, the Far UV (FUV) and Near UV (NUV) filters.  The FUV filter is
characterized by $\lambda_{eff}=1528$\AA, with a range of $1344-1786$\AA; 
the NUV filter has $\lambda_{eff}=2271$\AA, with a range of 
$1771-2831$\AA\ \citep{morr05}. 

We select QSO candidates from GALEX GR2, which includes 7077 GALEX AIS 
pointings covering approximately $8000$ square degrees on the
sky\footnote{http://galex.stsci.edu/GR2/} as well as $\sim350$ Medium Imaging
Survey (MIS) and $\sim100$ Deep Imaging Survey (DIS) 
pointings\footnote{http://galex.stsci.edu/GR2/}.  GR2 also
contains targets from the Nearby Galaxy Survey, which represent possible
contaminants in our catalog.
The GR2 pointings are broadly distributed over the sky, 
except regions within
$\sim$10$^{\circ}$ of the Galactic disk.  The AIS pointings, 
constituting the bulk 
of those available, have $1\sigma$ limiting magnitudes of 
$FUV=19.9$ and $NUV=20.8$ in the AB magnitude system.  The MIS 
pointings have $1\sigma$ limiting 
magnitudes of $FUV=22.6$ and $NUV=22.7$, and the DIS pointings have limiting 
magnitudes of $FUV=24.8$ and $NUV=24.4$
\citep{morr05}.  

We do not place any restrictions on the magnitude errors of
our candidates in order to allow greater freedom for follow-up studies to
set their own limits; the typical error on the NUV magnitudes is $\sim0.2$ mag,
while the typical error on the FUV magnitudes is $\sim0.3$ mag.  The 
distribution of errors in both bands has a long tail stretching to 
0.6 magnitudes.  Also, we assume that the internal processing performed to
generate the GR2 catalog of sources successfully eliminates artifacts and other
edge effects, and we do not consider contamination from these sources.

We match GALEX sources to the USNO-A2.0 catalog, which contains positions as 
well as $B$ and $R$ photometry for $\sim$$5\times10^{8}$ 
sources from the entire sky \citep{mone98}. 
The Palomar Optical Sky Survey (POSS) plates, 
from which the USNO-A2.0 catalog is derived,
are complete down to $V\approx21$ \citep{mone03}.  The R magnitude limit that we
impose on our candidates is 1.5 magnitudes brighter than this limit, 
so we should
approach the plate limits only for the
bluest objects in our input sample. We find empirically that objects bluer than
$B-R=-1$ are so extraordinarily rare as to be negligible.  

\begin{figure}
\epsscale{1.0}
\plotone{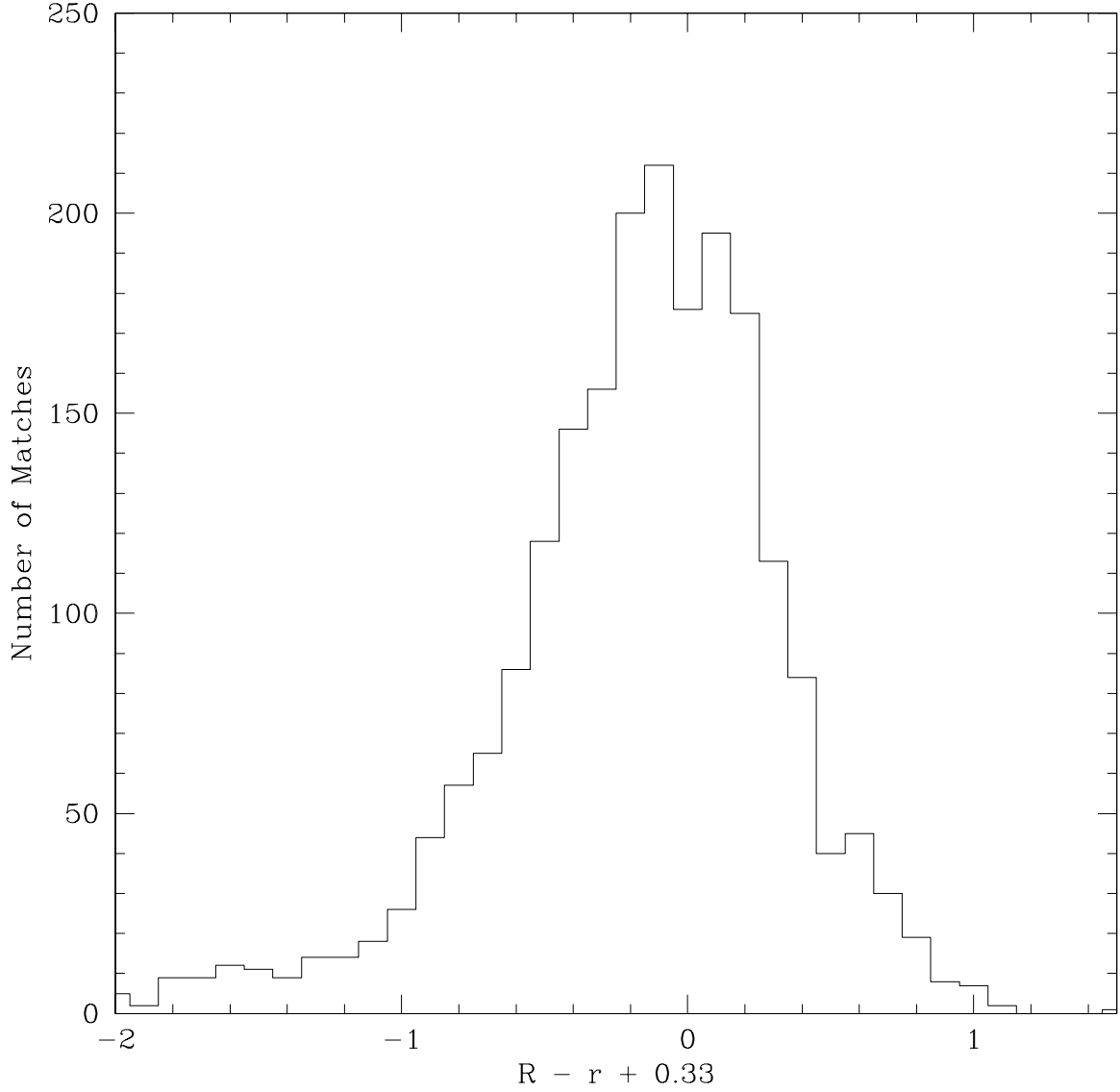}
\caption{Distribution of spectroscopically confirmed QSOs
appearing in both GR2 and DR4 and falling in 
$-0.05\leq{\it r-i}\leq0.05$.  The FWHM of this distribution is $\sim$0.8, 
so the 
$R$ magnitude uncertainty, given by $\sigma={\rm FWHM}/2.35$, is $\sim$0.35.
The additive factor of 0.33 on the x axis is an empirically estimated 
correction from $R$ to $r$.
\label{figRMags}}
\end{figure}

In order to examine the quality of the USNO $R$ band
photometry, we took from the GR2 sources confirmed to be QSOs by SDSS those
sources with $-0.05\leq{\it r-i}\leq0.05$ and plotted a histogram of $R-r$, 
shown in Figure \ref{figRMags}.  The additive term of 
$0.33$ on the x-axis is an (empirically) estimated correction 
to equate $R$ and $r$ band magnitudes.  We assume
that the dispersion about the mean is entirely due to the uncertainties in the 
USNO photometry, and find that the uncertainty in the USNO $R$ band photometry
is approximately 0.35 magnitudes.
This result is similar to the result of \citet{sali03}, who find the
USNO $R$ band photometry has an uncertainty of 0.25 magnitudes, but suggests
that the quality of the USNO photometry deteriorates at fainter magnitudes.

\begin{figure}
\epsscale{1.0}
\plotone{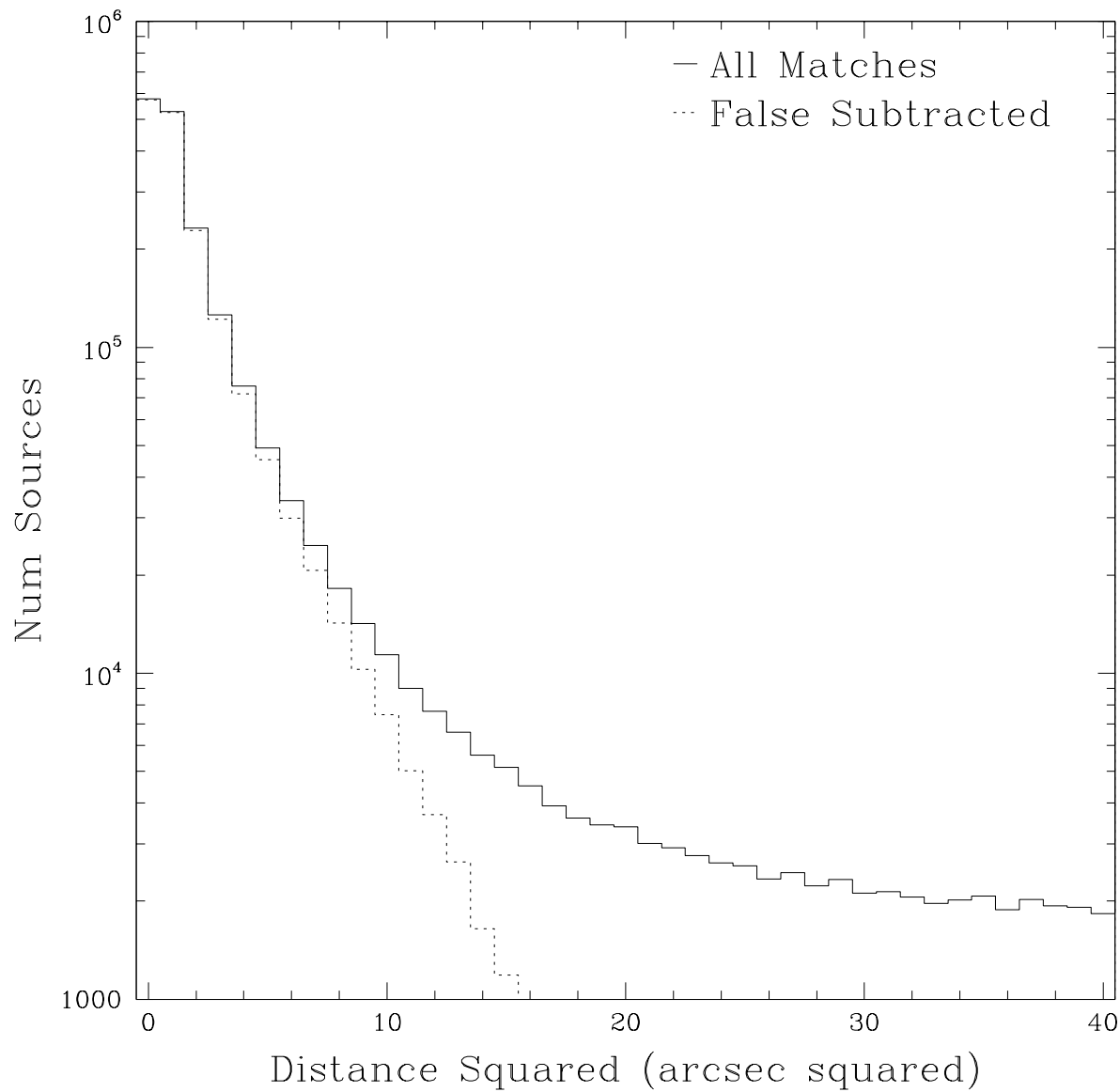}
\caption{
Distribution of squared distances between GR2 sources and the best 
(nearest) USNO counterparts ({\it solid}), and the same distribution with a 
2\% contamination due to random sources subtracted ({\it dashed}).  On these 
axes, a 2-d Gaussian projected into a 1-d distribution 
will appear as a straight line, the slope of which indicates a 1-d measurement
error of $\sigma=1''$.  Many matches between GR2 and USNO 
sources have match radii too large to fall in this plot.
\label{figLogDist}}
\end{figure}

The GALEX team has matched GR2 to SDSS DR4, and we
tune our selection criteria using this data set.
We match GALEX sources to USNO-A2.0 by submitting coordinates for all GALEX 
sources with detections in both FUV and NUV bandpasses to the VizieR 
search engine, requesting matches within $3''$ of each GALEX source.  In order
to fix a limiting match radius, we assume that the distribution of match 
distances should follow a 2-d Gaussian with contamination from uniformly 
distributed false matches.  On the axes in Figure \ref{figLogDist}, a 2-d
Gaussian projected into the radial direction yields a straight line; in order
to find the underlying Gaussian, we subtract a 2\% contamination from the 
distribution of match distances.  The false-subtracted distribution begins to 
diverge significantly from the total distribution near 10 arcseconds squared.
This suggests that we should choose a search radius near $3''$.

Using a $3''$ search radius, we 
translate 15\,000 GR2 sources $5'$ south and find 145 false matches, 
indicating a 1\% contamination by false matches at this radius.  This suggests
that $3''$ is an ideal search radius, including a lower fraction of 
contaminants than the total set of matches, but still including $3\sigma$ of
the GALEX astrometric uncertainty.  Since we expect from
the precision of our input photometry to have selection efficiency
less than 90\%, 1\% contamination of our candidates via mismatches is
acceptable.

From the GALEX-USNO matches selected by our search, we eliminate 
those with the USNO ``bad-magnitude'' flag set.  The resulting list of 
matches contains 1\,664\,229 total sources; this constitutes the input used to
develop our photometric selection criteria.  

The USNO PSF is much smaller than either the FUV or NUV PSF,
so we occasionally find multiple USNO counterparts
for a single GALEX source; we keep the USNO counterpart closest to the
GALEX position.  Because we restrict our final candidate set to fields with
$|b|\geq25^{\circ}$, the contamination of our catalog due to blended stars 
producing apparently non-stellar GALEX colors should be minimal.  
Our catalog of matched sources initially contains numerous double-reported 
GALEX sources,
since objects duplicated by repeated pointings at the same patch of sky have
not been merged (S. Salim, 2006, private communication).  We discuss the 
elimination of duplicated matches from our catalog in \S\ref{qsoCand}.  

\section{Photometric Limit Selection}\label{photo}
We seek to develop criteria, using GALEX GR2 and USNO-A2.0 photometry, that
select QSO candidates with a high probability of being real and
retain as
many candidates as possible.  The high completeness \citep{rich02}
and excellent photometry of the SDSS quasars provide
a superb calibration sample with which to tune these criteria.

We begin by identifying the subset of GALEX-USNO sources for
which SDSS DR4 is essentially complete with regard to QSO selection.
We demand that the source appears in GALEX's set of SDSS 
matches and has $i\leq19.1$, since SDSS
systematically searched for quasars only down to this limit.  Also,
we require that there be at least one object within $7\farcm{6}$
of the source with an SDSS DR4 spectrum.  We select this radius by assuming
an average density of 100 spectroscopic targets per square degree, and then 
requiring that there should be an average of five targets within our 
search radius.  (We find by direct search
that only 0.2\% of spectroscopically observed targets have no additional
target within the search radius.)

\begin{figure}
\epsscale{0.8}
\plotone{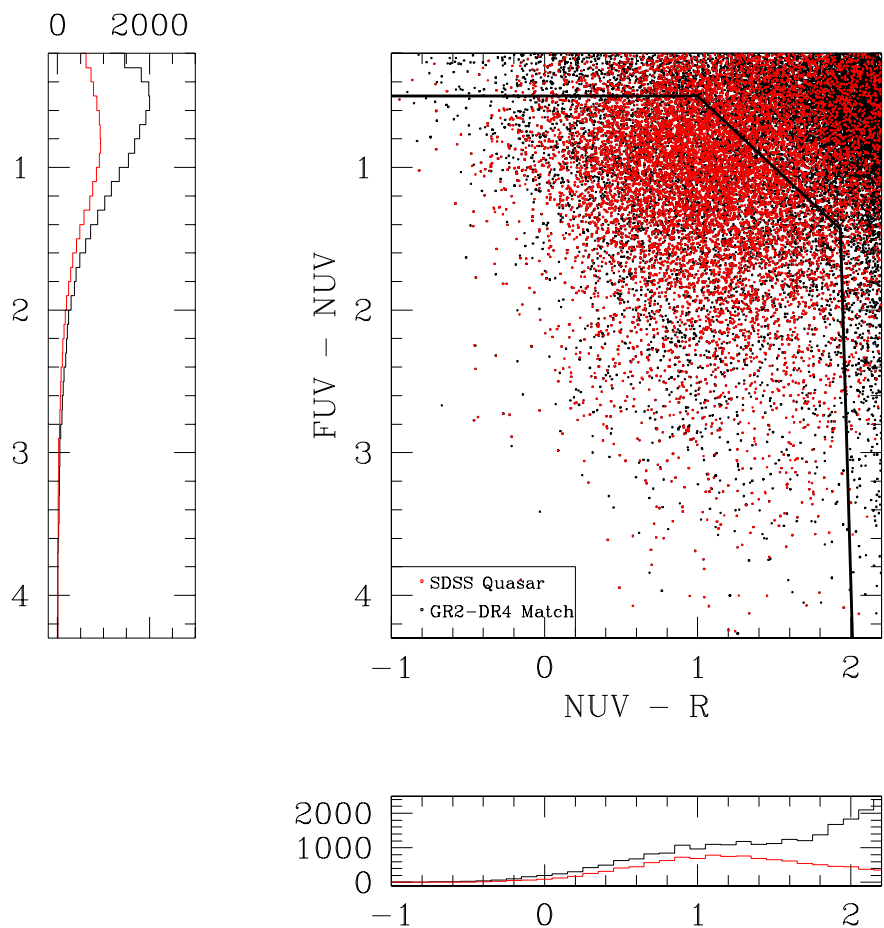}
\caption{Selection criteria ({\it bold lines})
for QSO candidates on a color-color
diagram derived from GALEX ($FUV$ and $NUV$) and USNO-A2.0 ($R$)
photometry.  SDSS quasars are in {\it red}, and all other sources
are in {\it black}.  All displayed sources have $i\leq19.1$.  The histograms
at bottom and at left are projections of the displayed sources onto the relevant
color axes.  The black histograms show the total number of sources, and the
red histograms show the number of SDSS QSOs.  The numerous sources
above and to the right of the displayed color range are excluded from the
histograms.
\label{galexQsoColors}}
\end{figure}

We display all sources satisfying these three criteria on an
$FUV-NUV$ vs.\ $NUV-R$ (GALEX-USNO) color-color diagram, using red to
mark SDSS QSOs and black for all other objects.  We
experiment with various polygons in the GALEX-USNO color-color plane
to maximize the number genuine QSOs while minimizing the
number of contaminants.  Figure \ref{galexQsoColors} shows our adopted
polygon and the distributions of QSOs and non-QSOs within it, as well as
just outside the selection area.

We place an additional constraint on the $B-R$ color of our candidates,
in effect placing our selection criteria in a three-dimensional color space,
in analogy with the procedure adopted by \citet{rich02} for SDSS.
This additional cut provides only limited discrimination between QSOs and
contaminants, due to the large uncertainties in USNO photometry, 
but it eliminates some outlying stars.
We list our adopted color selection criteria in Table \ref{tblBounds}.

In addition to our color criteria, we find that the number of 
GALEX-USNO matches drops rapidly for $R\geq 19.6$, and we therefore 
limit our catalog to $R\leq 19.5$.
Finally, by comparing the Galactic-latitude distribution of
our QSO candidates to QSOs found in the Veron catalog \citep{vero06},
we find that we have very little sensitivity for $|b|<25^\circ$ and
therefore do not search for quasars below this limit.  We believe this
low sensitivity is due to heavy extinction in the $FUV$-band.

\section{QSO Candidates}\label{qsoCand}
After performing our photometric selection, we 
eliminate the duplicate candidates that entered the catalog through 
multiple GALEX exposures by looking for consecutive objects separated by less
than $1''$.  We selected this limit because \citet{sieb05} indicate that 
GALEX astrometry is accurate to
within $1''$.  We tested this radius by doubling it to $2''$ and found 
that only eight additional GALEX objects were eliminated, so we deem a $1''$
limiting radius sufficient for our purposes.

All candidates we determine to be associated with the same 
physical object are examined
to determine whether the UV magnitudes changed between observations.
Those candidates whose $FUV$ and $NUV$ magnitudes both changed by more than
$1\sigma$ are flagged as variable.  For these candidates, we compute average
magnitudes using all available observations and record the averages in 
our catalog.  For candidates that do not exhibit
UV variability, we use the magnitude with the lowest listed uncertainty.
Our initial sample of QSO candidates, selected as described in \S\ref{photo},
includes 43\,345 GALEX sources 
down to a limiting magnitude of $R=19.5$.  After eliminating duplicate 
GALEX sources, we retain 36\,120 candidates, which comprise our final catalog.

\subsection{Candidate Probability Assessment}\label{probAss}
We assess the probability that each candidate is a QSO by assuming 
that the ratio of QSOs to non-QSOs at fixed USNO-GALEX colors is 
independent of position on the sky, and only weakly dependent on apparent 
magnitude.  In particular, we assume that the probability for a candidate with
$i\leq19.1$ (for which we have substantially more data) is an accurate predictor
of the probability for a fainter candidate with similar colors.

\begin{figure}
\epsscale{0.8}
\plotone{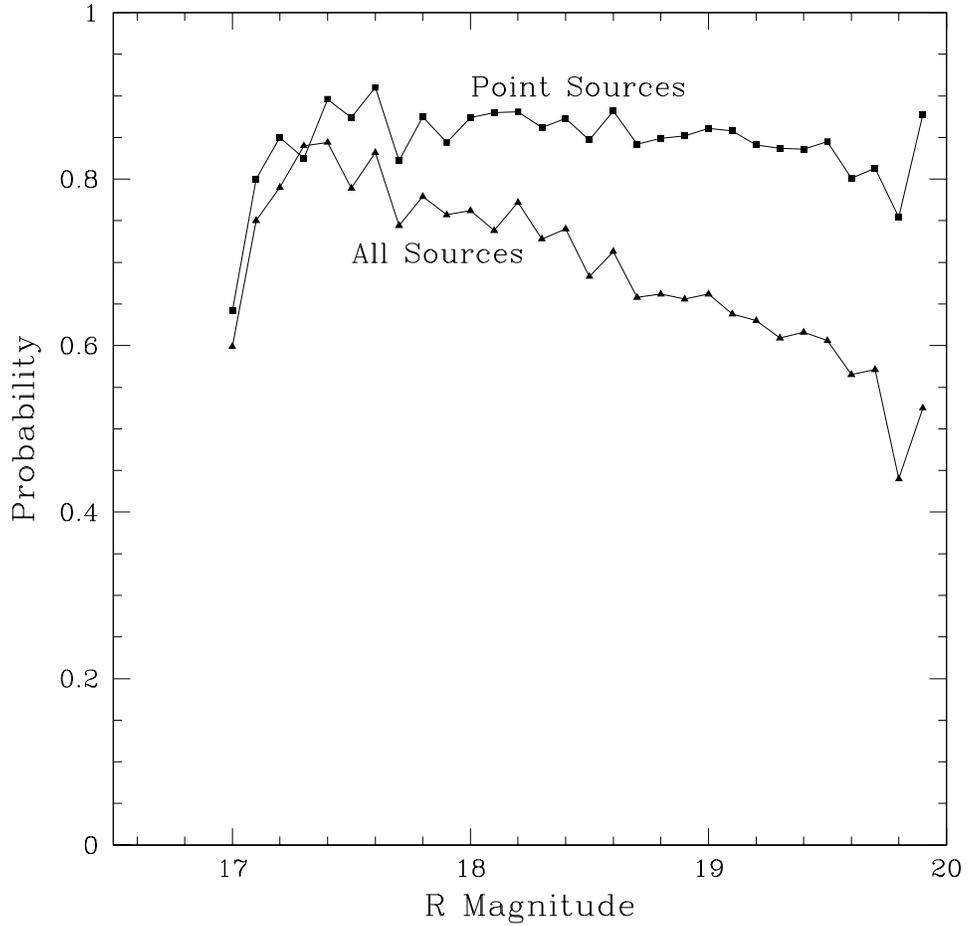}
\caption{Probability that a candidate with a given $R$ magnitude will fall
inside our ``SDSS QSO Selection Area'' in Fig.~\ref{sdssQsoColors},
both for all sources ({\it triangles}) and SDSS point sources 
({\it squares}).  
The probabilities shown for 
$R=17.0$ include all candidates with $R\leq17.0$.
\label{rFraction}}
\end{figure}

\begin{figure}
\epsscale{0.8}
\plotone{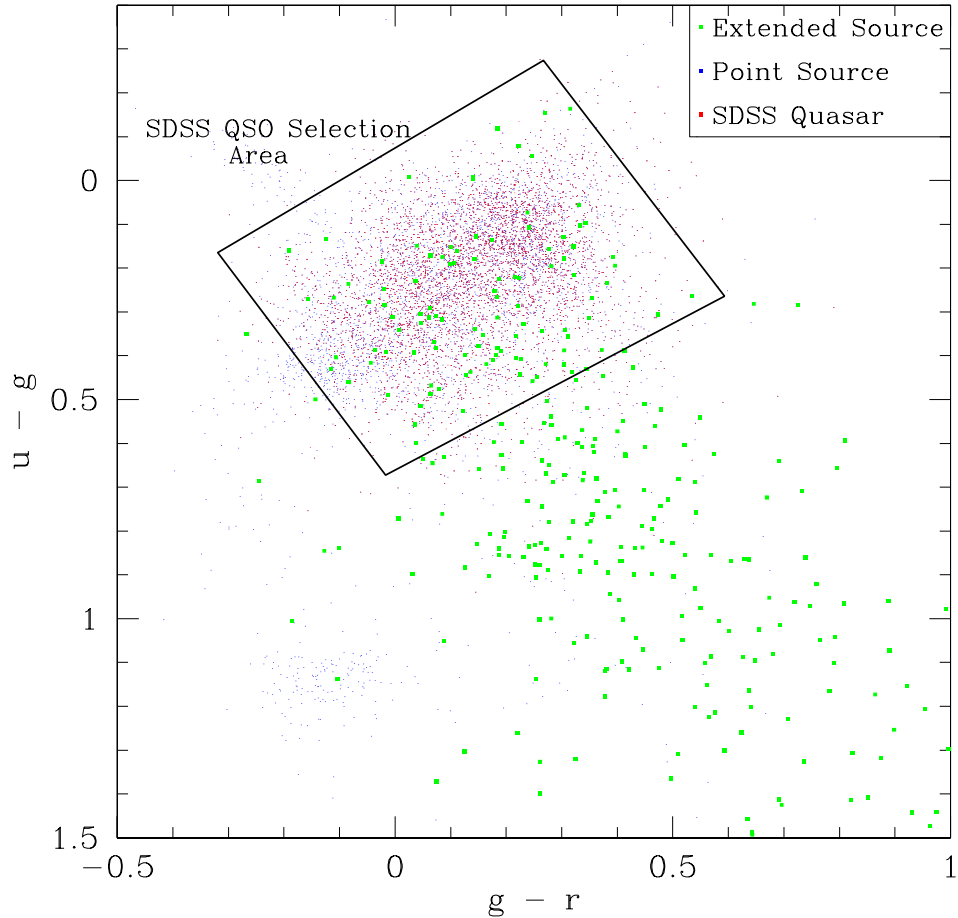}
\caption{Distribution of candidates in the SDSS $ugr$ color-color space with  
QSOs ({\it red}), other SDSS point sources ({\it blue}),
and SDSS extended sources ({\it green}).  The heavy line
indicates the boundary of our ``SDSS
QSO Selection Area,'' used in the calculation of QSO probabilities.  
(See \S\ref{probAss}.)  This figure
includes only candidates with $i\leq19.1$; the corresponding diagram for sources
with $i>19.1$ has a similar overall shape, but there are more 
extended sources and significantly fewer SDSS quasars at 
fainter magnitudes.
\label{sdssQsoColors}}
\end{figure}

\citet{bian06} indicate that the fraction of QSOs at fainter
magnitudes is enhanced with respect to stars, so this is a potential source of
bias in our probability calculation.  If we were strongly affected by this bias,
we would under-predict the probability for faint candidates to be QSOs and
over-predict the probability for bright candidates to be QSOs.  We discuss
this implication further in \S\ref{nbcKde}.
Since Figure \ref{rFraction} shows that the fraction of 
selected point sources
with SDSS colors similar to the SDSS QSOs is approximately constant 
with R magnitude, we believe that the effects on our probability
calculation due to this bias are modest.
The variation is stronger if extended sources are included as well, which is
unsurprising if one considers the fact that QSO and galaxy colors show
significant overlap even in SDSS colors.  (See Fig.~\ref{sdssQsoColors}.)
Nevertheless, the
trend is still sufficiently regular that it should be correctable by the 
procedure we discuss below.

Most SDSS quasars occupy a relatively small
region in $ugr$ color space, which we call the ``SDSS QSO Selection 
Area''\footnote{The use of this name does not indicate any 
specific correlation with
the SDSS spectroscopic target selection algorithm; it is simply a term of
convenience.  For details on the procedure SDSS uses to select quasar
candidates for spectroscopy, see \citet{rich02}.}.  Figure \ref{sdssQsoColors}
shows where various types of sources fall within SDSS $ugr$ color space,
and indicates the boundaries of the region we call the ``SDSS QSO Selection
Area.''
Taking advantage of this correlation, we scale the probability according to the 
$R$-magnitude of the candidate by accounting for the number of candidates that 
fall into our ``SDSS QSO Selection Area'' as a function of $R$,
\begin{equation}\label{eqProb}
P\left(R,FUV,NUV\right) = P_{0}\left(FUV-NUV,NUV-R\right)\left[\frac{N_{
\rm qso\,area}}{N_{\rm all}}\right]_{R}\left[\frac{N_{\rm qso\,area}}{N_{
\rm all}}\right]^{-1}_{i\leq19.1}
\end{equation}
where $N_{\rm qso\,area}$ is the number of sources that fall within the
region defined by the heavy lines in Figure \ref{sdssQsoColors}, and
$P_{0}\left(FUV-NUV,NUV-R\right)$ is determined by dividing 
the number of SDSS QSOs within a circle of radius $0.05$
mag centered at $(FUV-NUV,NUV-R)$ by the total number of point sources in
the same circle.  The number of sources indicated by $N_{{\rm all}}$ varies 
according
to the $R$ parameter, and gives the total number of sources of magnitude $R$;
$N_{{\rm all}}$ for $i\leq19.1$ is the total number of sources with $i\leq19.1$.

The first factor in equation (\ref{eqProb}) would give the probability that a
candidate is a QSO if it were (somehow) known {\it a priori} to have
$i\leq19.1$.  However, for many candidates there are no SDSS data.  
The second factor accounts for the fraction of sources at fixed $R$
that fall into the $ugr$ color-color region including most SDSS
quasars.  Its dependence on $R$, both for all sources and for SDSS point
sources only, is shown in Figure \ref{rFraction}.
Finally, the third factor is the correction accounting 
for the fact
that not all SDSS quasars with $i\leq19.1$ actually fall into the 
``SDSS QSO Selection Area,''
and that this missing fraction is already included when calculating
$P_{0}\left(FUV-NUV,NUV-R\right)$.

For the candidates in our catalog, we calculate separate probabilities
that the candidate will be an extended source, and therefore not a QSO
(see \S\ref{photo}), and that
a given candidate will be a QSO, assuming that it is a point source.  The
extended-source 
probabilities are computed by using the density of SDSS extended sources 
relative to SDSS point sources in proximity to the candidate in USNO-GALEX 
color-color space,
\begin{equation}\label{exProb}
P_{ext}=1-\left[P_{i\leq19.1}P_{ps,i\leq19.1}+\left(1-P_{i\leq19.1}\right)P_{ps,i>19.1}\right]
\end{equation}
where $P_{i\leq19.1}$ is the probability that a candidate with a given 
$R$-magnitude will have $i\leq19.1$, while $P_{ps,i\leq19.1}$ and 
$P_{ps,i>19.1}$ are the conditional probabilities that a candidate with
a given color will be a point source for $i\leq19.1$ and $i>19.1$, 
respectively.  The conditional probabilities employed here are calculated 
in the same way as $P_{0}\left(FUV-NUV,NUV-R\right)$
from equation (\ref{eqProb}).

\begin{figure}
\epsscale{0.8}
\plotone{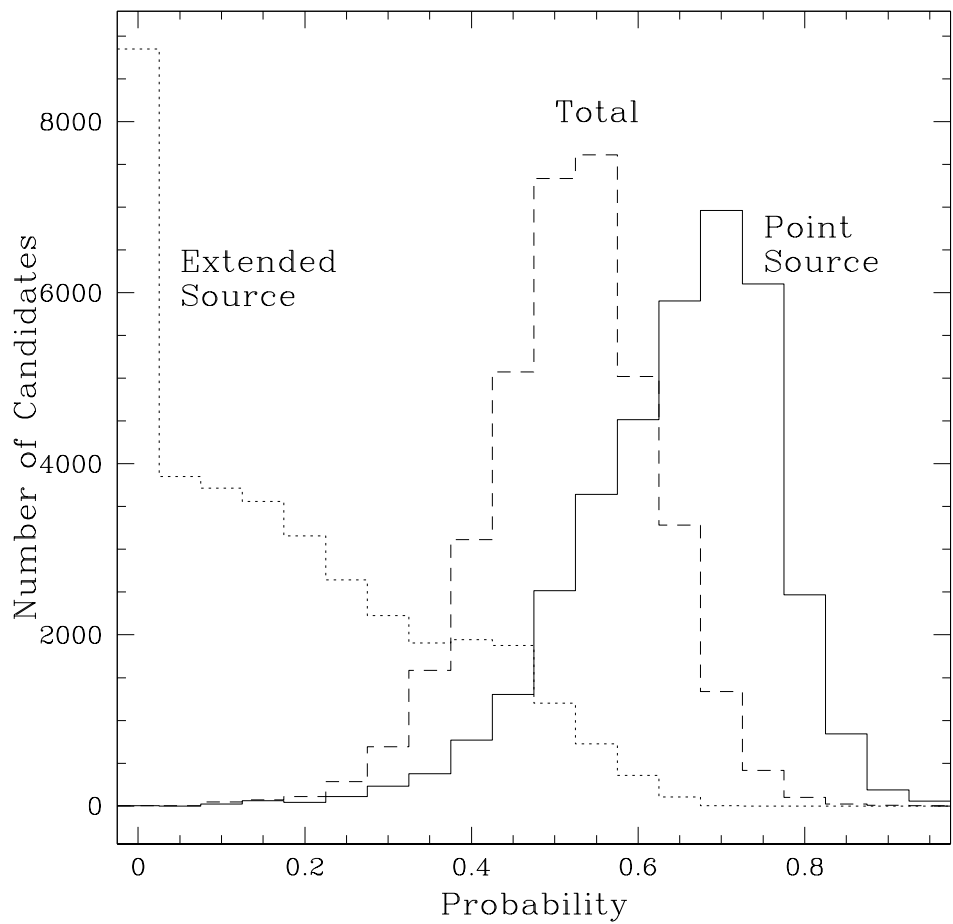}
\caption{Distribution of calculated probabilities that candidates are
actually quasars.  The {\it long-dashed} histogram gives the 
distribution of total probabilities that candidates will turn out to be QSOs; 
the {\it short-dashed} histogram shows the distribution of 
probabilities that
candidates would be identified as extended sources in a short imaging exposure,
and the {\it solid} histogram shows the distribution of 
conditional probabilities assuming that all candidates
were subsequently identified as point sources. 
\label{probDist}}
\end{figure}

The total probability that a given source is a QSO is obtained by 
multiplying the probability that a
point source with given colors is a QSO ($P_{qso,ps}$)
by the probability that the candidate is a point source, i.e.,
\begin{equation}\label{totalProb}
P_{qso}=P_{qso,ps}\left(1-P_{ext}\right)
\end{equation}
The distribution of total probabilities that we calculate using equation
(\ref{totalProb}) is labeled ``Total'' in Figure \ref{probDist}.  

Upon integrating the total probabilities, we find that $\sim20\,200$
of our candidates should be genuine QSOs, yielding a selection efficiency of
$52\%$ for our catalog as a whole.
This efficiency is reasonably good, in that only half the time of a 
spectroscopic follow-up study would be spent taking spectra of non-QSOs.  The 
efficiency could be further improved by ``pre-screening'' the 
candidates with snapshot images to eliminate extended sources.  
This would require only $\sim$1 minute exposures on a 1m telescope, and
would improve the overall efficiency to $65\%$.

It should be noted that, while we have matched many of our candidates to various
existing catalogs, we have not taken any such matches into account in our 
probability calculation, except insomuch as we have used SDSS quasars to 
determine $P_{0}\left(FUV-NUV,NUV-R\right)$
and SDSS photometry to determine the conditional probabilities
$P_{ps,i\leq19.1}$ and $P_{ps,i>19.1}$.  In particular, we have not
elected to use the morphology indicator available from USNO-B to
modify the probability of a candidate being extended or to eliminate candidates
with apparently extended morphologies.  
See \S\ref{usnoB}, below, for more information on our matching to USNO-B
and a discussion of the USNO-B fields that we have included in our catalog.

Finally, our examination of matches between our catalog and sets of
spectroscopically confirmed QSOs indicates that the probabilities we calculate
provide a good statistical description of the candidates, but of course do not
predict whether any given candidate will be a QSO.  

\subsection{Matching With USNO-B}\label{usnoB}
It should be possible to reduce the fraction of telescope time
spent observing poor candidates by including additional information
available from the USNO-B catalog.  Specifically, the contamination by 
foreground stars could be reduced by including the measured proper motions.
In addition, the time spent observing extended sources, unlikely to be QSOs,
could in principle be reduced by including star-galaxy separators.  
Thus, we match our catalog of
candidates to USNO-B employing the same procedure described in \S\ref{input}
above, except allowing a $5''$ search radius.  We choose to allow an expanded
search radius because the significantly smaller number of input objects
reduces the number of spurious matches to an acceptable level.

Eliminating candidates with proper motions greater 
than twice the inherent uncertainty in the USNO-B proper motions 
leaves $27\,752$ 
candidates out of $36\,089$ with matches in the USNO-B catalog.  (See 
\citealt{mone03} and \citealt{goul03} for discussion of the 
astrometric precision in USNO-B.)
Application of this cut would reduce the number of SDSS quasars with proper 
motions measured by USNO-B from $5153$ to $4566$, i.e. by
$11\%$, compared to a $23\%$ reduction in the total number of candidates.
While the quoted proper motion cut would improve the fraction of genuine
QSOs in the catalog, we have not eliminated the candidates with measurable
proper motions from our catalog because a significant fraction of genuine QSOs
would also be eliminated.  We elect instead to list the proper motions and allow
individual users to decide how to apply a proper motion cut.

\begin{figure}
\epsscale{0.8}
\plotone{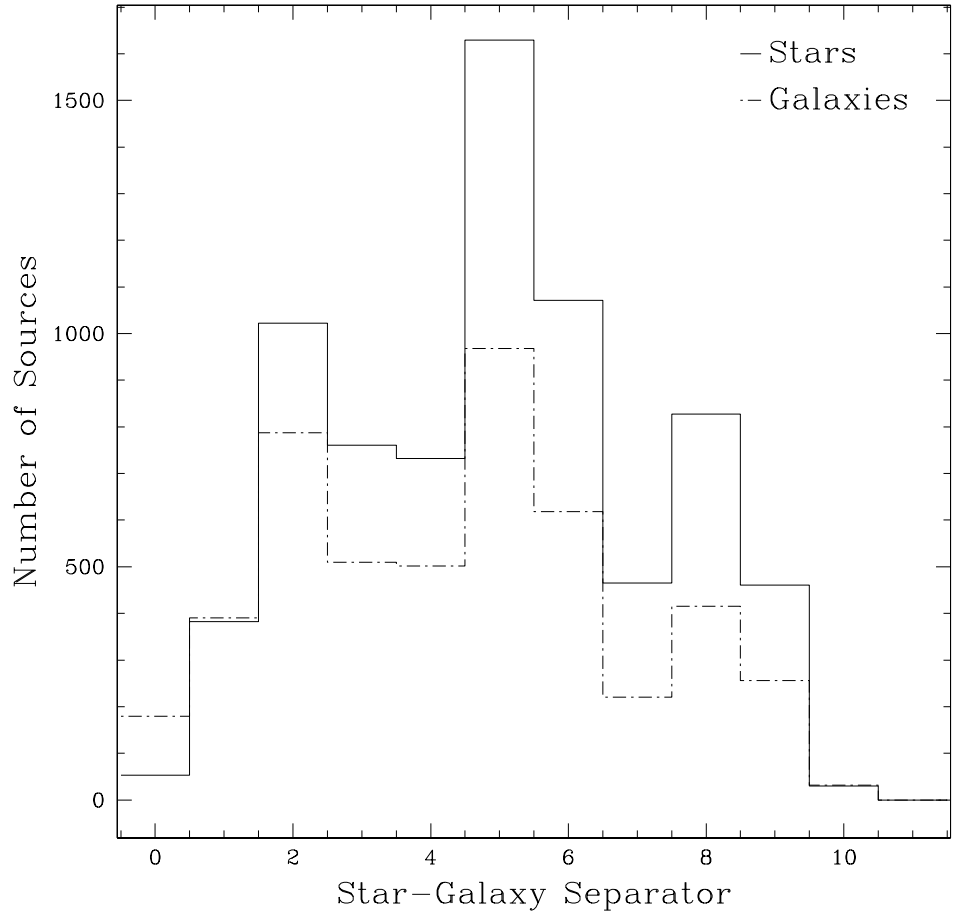}
\caption{Distributions of USNO-B
 $R$ band star-galaxy separators for candidates with
known morphologies from SDSS.  The {\it solid} histogram is for morphological
stars, and the {\it dot-dashed} histogram is for morphological galaxies.  A
star-galaxy separator between 8-11 is considered a strong indicator of star-like
morphology in the USNO-B catalog, while a value between 0-3 is considered 
indicative of galaxy-like morphology.
\label{sgCompare}}
\end{figure}

We similarly choose not to eliminate candidates that do not appear point-like on
the POSS plates.  This is because there appears to be only
a weak correlation between the star-galaxy indicators from USNO-B and the
morphologies reported by SDSS.  In Figure \ref{sgCompare} the distribution of
star-galaxy separators in the $R$ band is plotted for candidates with known SDSS
morphologies.  While there is a clear trend for galaxies to be less PSF-like
than stars, the significant overlap between the two distributions belies
the ability of the USNO star-galaxy separator to distinguish galaxies from
stars at a level sufficient to improve our selection efficiency.  As a result,
we include the calculated probability for a candidate to be extended
and the star-galaxy separators for both $B$ and $R$ bands in our catalog,
but do not employ them to eliminate candidates.

\subsection{Comparison With Sloan Digital Sky Survey}\label{sdss}
Of the $36\,120$ QSO candidates selected by our algorithm, 
$18\,284$ fall within the area covered by SDSS DR4, and 
$5187$ are SDSS quasars.  There are a total of $5969$ 
candidates that have SDSS spectra, so at least $782$ candidates 
in DR4 are not SDSS quasars.  Of the DR4 sources without SDSS spectra, $10\,119$
have $i>19.1$, and so were unlikely to have been 
selected for SDSS spectroscopy \citep{rich02}.  Of the remaining candidates, 
$2144$ fall more than $7\farcm{6}$ from the nearest DR4 spectroscopic target,
indicating that they lie in regions where spectroscopy is not yet available.
(See \S\ref{photo}.)

There are a total of $19\,116$ SDSS QSOs with $i\leq19.1$ having
counterparts in GR2, of which $4277$ are identified as candidates by our
criteria.  This yields a total completeness of $22\%$ for our
selection criteria with respect to the SDSS quasars.  If we instead
consider the number of SDSS quasars with USNO-A2.0 counterparts and detections
in both the FUV and NUV bands, we find 11\,321, yielding a completeness of 
$\sim38\%$.  Of the remaining 7795 sources, most (6935) were excluded from
our search because they had no $FUV$ detections.  

\citet{bian05} develop a selection algorithm
qualitatively similar to ours, and have sensitivity only to $z\approx2$.
The large number of SDSS QSOs lacking $FUV$ detections is due, at
least in part, to the continuum flux of the higher-$z$ sources being redshifted
out of the sensitivity range of the $FUV$ bandpass.  As a result, we will
preferentially miss high-$z$ QSOs.  
Most SDSS QSOs with extended morphologies appear redder 
than our selection boundary; this suggests that we also preferentially
miss nearby, and thus low luminosity, SDSS QSOs.
We know that there is also a significant
population of point-like SDSS QSOs redder than the 
selection boundaries we have imposed.
Some of these may be strongly reddened by intervening dust, 
suggesting that we will not 
be sensitive to highly absorbed QSOs.  This is a natural consequence of
searching for UV-excess QSOs; an IR survey like
that of \citet{bark01} is preferred to find strongly absorbed QSOs.

\begin{figure}
\epsscale{0.8}
\plotone{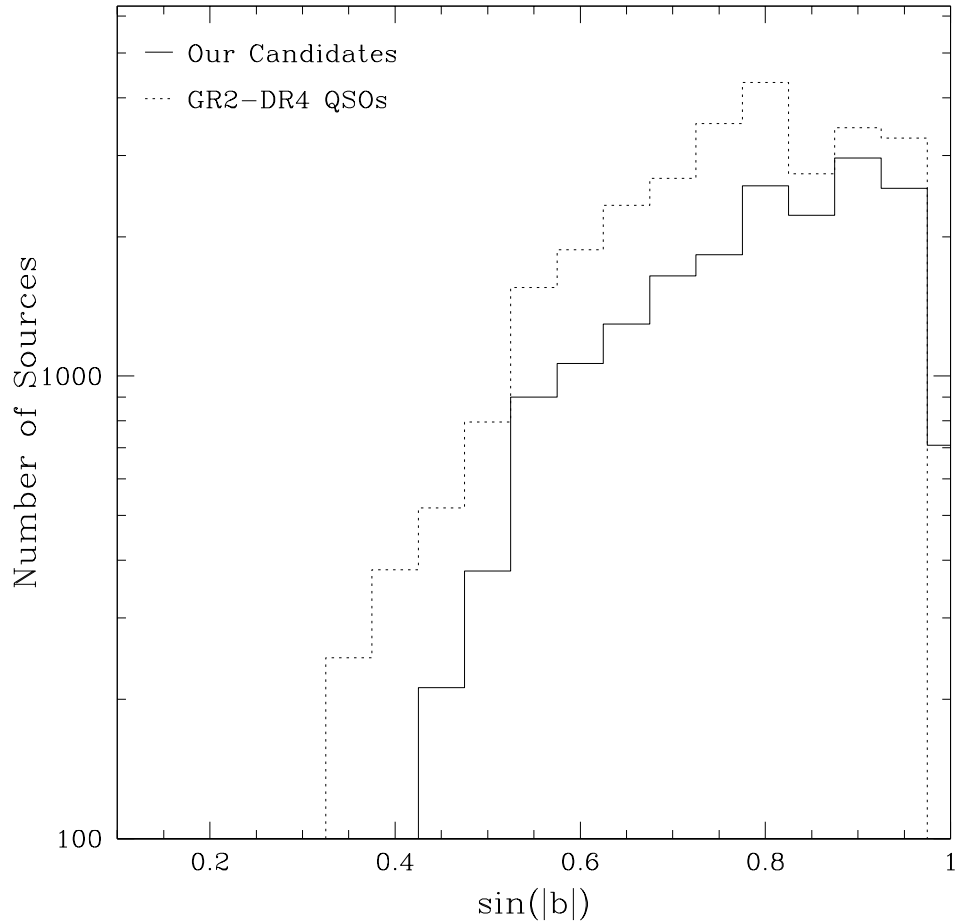}
\caption{Distribution of Galactic latitudes for candidates with
DR4 counterparts {\it solid} and known QSOs from 
GR2-DR4 match ({\it short-dashed}).  The distribution of candidate latitudes
closely follows that of previously known GR2 QSOs, indicating that the selection
procedure we have developed does not preferentially select candidates of 
different Galactic latitudes.  The significant drop in the number of candidates
near $sin(|b|)=0.4$ is due to our latitude cut at 25$^{\circ}$.
\label{figCompLatitude}}
\end{figure}

In Figure \ref{figCompLatitude}, we compare the Galactic latitude distribution
of DR4 QSOs with GR2 counterparts to that of the subset of our candidates with
DR4 counterparts.  
Our candidates and the DR4 QSOs with GR2
counterparts follow roughly similar latitude distributions, indicating that
our assumption in \S\ref{probAss}, that the probability for a given candidate
to be a genuine QSO is independent of position on the sky, will provide a valid
representation of candidate probabilities.  This is not 
surprising, but its verification helps rule out strong influence by
dust reddening in the Galactic disk.  It is also worth noting that our
selection procedure does fractionally better at very high Galactic latitude,
due to larger sky coverage at very high latitude by GR2.

\subsubsection{Candidates from the Kernel Density Estimate}\label{nbcKde}
\citet{rich04} select a set of UV-excess quasar candidates using SDSS 
photometry from DR1.  The
selection criteria they employ were subsequently applied to DR3 sources and
the results published on the SDSS 
website\footnote{http://sdss.ncsa.uiuc.edu/qso/nbckde/}.  We have compared
our candidate set to theirs and were thereby able to obtain another
measure for the selection efficiency of our algorithm, independent of the 
$i\leq19.1$ limit imposed by SDSS for QSO candidates.

The {\it kernel density estimate} (KDE) candidate selection algorithm shows a
95.0\% selection efficiency after correction for magnitude bias \citep{rich04}.
We find 9119 unique candidates with
matches in the KDE selected catalog, from a total of 15\,860 
candidates with DR3 matches.
Assuming the quoted 95.0\% efficiency of the \citet{rich04} candidate set, we 
estimate an efficiency of 55\% for our selection criteria.  This is essentially
the same as the 52\% we calculated in \S\ref{probAss} above.  The
consistency of these two independent estimators indicates that our 
probability calculations are performing reasonably well.

As discussed in \S\ref{probAss}, the influence of a magnitude
dependent bias would cause our probability assessment to differ from the 
``true'' probability at both the bright and faint ends of our catalog.  The
much larger number of faint candidates relative to bright ones would result in
a significant under-prediction in the number of total ``good'' candidates in
our catalog, if we were strongly influenced by this effect.
The confirmation of our probability calculation using the KDE-selected sample,
which does not rely on the assumption that the probability for a bright
candidate to be a QSO predicts the probability for a fainter object 
of similar colors to be a QSO,
suggests that our probability calculation is not adversely impacted by 
the magnitude dependent variation in the fraction of QSOs with respect to
hot stars, reported by \citet{bian06}.  This in turn implies that we
successfully avoid the influence of magnitude dependent effects.

Our avoidance of such effects
arises because we deliberately avoid selecting candidates from
the stellar locus, insofar as is possible, in conjunction with our use 
of a magnitude dependent
correction for the contaminant fraction.  In addition, we do not include
candidates significantly fainter than the limit for SDSS spectroscopy,
thereby limiting the potential influence of 
magnitude dependent effects.  It is, of course, possible that
the dominant fraction of extended sources among the contaminants masks the
effects of a magnitude dependent bias.  The approximately constant fraction
of point-like candidates with QSO colors, apparent in Figure~\ref{rFraction},
belies this interpretation, but we cannot absolutely exclude it 
without the results of an extensive spectroscopic follow-up
beyond $19^{\rm th}$ magnitude.

\subsection{Comparison With Other Catalogs}\label{comp}
In addition to SDSS, we compare our candidates to sources in the 2 Micron 
All-Sky Survey (2MASS), the 
Faint Images of the Radio Sky at Twenty cm (FIRST)
survey, the XMM-Newton Slewing Survey (XMMSL1) \citep{frey06} and
the most recent Veron catalog of known QSOs \citep{vero06}.  We 
match our candidates to the Veron QSOs and 2MASS point sources via the
VizieR search engine, requiring that the 
distance from the matched source to the candidate be less than $5''$.
We find $5889$ candidates with counterparts in the Veron catalog out of $9013$ 
Veron QSOs with GR2 detections in both the FUV and NUV bandpasses.
This corresponds to a completeness of 65\%.  The large difference between
our completeness with respect to Veron and SDSS QSOs is somewhat surprising, 
given that the \citet{vero06} catalog includes DR4 QSOs.  This higher 
completeness arises because we find a
lower number of total matches between GR2 sources and Veron QSOs than GALEX
found between GR2 and DR4.
\citet{sieb05} indicate that a $6''$ search radius was used in the original
match between GALEX sources and SDSS DR1; however, \citet{bian06} indicate that
the match performed between GR1 and DR3 used a $4''$ search radius, which is
smaller than ours.  Depending on which search radius was employed for the
matching between GR2 and DR4, it may be trivial to understand why we find fewer
matches between GR2 and the \citet{vero06} catalog, or it may prove quite
difficult to explain.  Because we have been unable to determine the search
radius used, we do not speculate further.

We also find $4226$ candidates with 2MASS counterparts from a total of 
$6879$ objects in the 2MASS All-Sky Catalog of Point Sources having GALEX
detections in both the FUV and NUV bandpasses \citep{cutr03}.  
In their discussion of photometric selection of obscured QSO
candidates using 2MASS colors, \citet{bark01} indicate that $B-J<2$ is  
characteristic of most optically-selected QSOs.  While not necessarily 
indicating whether the selected candidates are genuine, the fact that $90\%$ of
our candidates have $B-J<2$ is reassuring.  
Also useful is the fact that, because these 2MASS sources are members of the 
point source catalog, they are less likely to be galaxies than candidates in the
general population.  Specifically, we find that $26.6\%$ of candidates with
SDSS counterparts exhibit extended morphology in SDSS photometry, while only 
$6.8\%$ of candidates with both SDSS and 2MASS counterparts are extended.
However, this strong improvement is due largely to the relatively bright
limiting magnitude of the 2MASS survey.
If we restrict our examination to candidates with $R\leq17.7$, 
beyond which point the number of candidates with 2MASS 
counterparts as a function of magnitude begins to decrease,
we find that $6.45\%$ of candidates with SDSS counterparts are 
extended, compared to $2.75\%$ of those with both 2MASS and SDSS counterparts. 
Thus, bright candidates with 2MASS counterparts are somewhat less likely to
exhibit extended morphology than other candidates, and would therefore yield a
slight improvement in observational efficiency compared to the catalog as a
whole.

The procedure we use to match our candidates with the FIRST catalog is 
similar, with the exception that we do not use the VizieR search engine but
the web-based search tool provided by the FIRST
collaboration\footnote{http://sundog.stsci.edu/cgi-bin/searchfirst}.
Again using a limiting search radius of $5''$, we find that 
$720$ of our candidates have counterparts in the FIRST catalog.

To match our candidates with the objects contained in XMMSL1, we
acquire the complete catalog from the XMM-Newton Science Archive website
and extract the columns of interest.  We eliminate any source with 
a warning flag and search the remaining objects for positions
within $2\sigma$ of our candidates.  We take this approach,
rather than using a fixed search radius, because of the large variation in
astrometric uncertainty from one XMMSL1 object to the next. 
From the $2692$ sources in XMMSL1 \citep{frey06}, we find $20$ that 
match our QSO candidates.  The total probability of these matches predict that
ten should be genuine QSOs; ten of these candidates appear in \citet{vero06},
with little information for most of the others.  One of 
candidates identified as a QSO in \citet{vero06}
(USNO 1275-07898737) is unusual; it has an SDSS spectrum but cannot be 
classified by the automated pipeline.

In addition to our matching, we check the likelihood that spurious matches 
between our candidates and the various catalogs will cause significant 
contamination.  For FIRST, Veron and
2MASS, we translate 15\,000 candidates south by $5'$ and repeat our matching
procedure.  For XMMSL1 we elect to translate the entire catalog of sources
rather than doing a simple representative test because of the 
very small number of matches from XMMSL1.  We find zero spurious matches in all
four tests, indicating an exceedingly low probability for spurious matches 
between our candidates and the supplementary catalogs.  

\begin{figure}
\epsscale{0.9}
\plotone{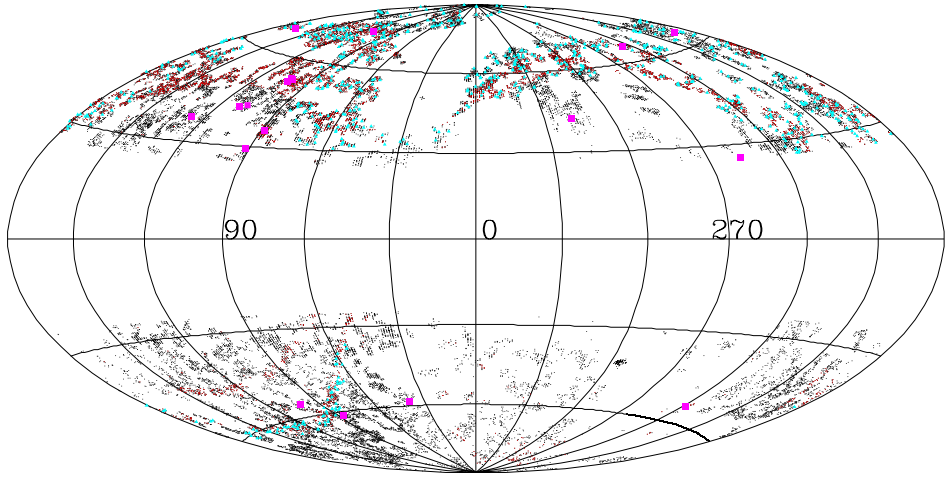}
\caption{Aitoff Projection of candidates in Galactic coordinates, including
5889 known QSOs with counterparts in Veron (mostly SDSS) ({\it red}), 
702 candidates with FIRST counterparts ({\it cyan}) and 20 counterparts in XMMSL1 
({\it magenta}).  All remaining candidates are shown in {\it black}.
\label{skyMatches}}
\end{figure}

See Figure \ref{skyMatches} for Galactic positions 
of our candidates, including indicators of matches to the Veron, 
FIRST and XMMSL1 catalogs.

\section{Catalog Description}\label{descr}
The catalog of candidates, which is organized by Right Ascension 
and appears in Table \ref{tblCand}, includes $36\,120$ candidates, after
duplicated GALEX observations have been eliminated.  It contains $21$ 
fields, including sky coordinates and identifiers from both the USNO-A2.0 and 
GALEX catalogs.  Also 
included are two flags classifying the matches we have made to other 
catalogs.  The first is the character flag indicating matches with Veron, 
2MASS, XMMSL1 and FIRST.
This flag consists of one or more characters that are
added for matched candidates; `V' indicating a match to \citet{vero06}, `M'
to 2MASS, `X' to XMMSL1 and `F' to FIRST.  
In addition to these character flags, there is an integer flag indicating
the nature of any SDSS counterpart.  The allowed values of
this flag and their meanings are summarized in Table \ref{tblSdssFlag}.

As discussed in \S\ref{usnoB}, we have included proper motions and mean
star-galaxy separators from USNO-B
for the great majority of our candidates.  Proper motions
are listed in the catalog in
mas ${\rm yr}^{-1}$, as they are in the USNO-B catalog.  Star-galaxy separators
take a value from 0 to 11, with 11 indicating sources that are very similar to
the POSS PSF and 0 those that are very dissimilar.  USNO-B lists a star-galaxy
separator for each detection on the POSS plates; we have averaged the
star-galaxy separators for each magnitude where observations at multiple epochs
are available.  \citet{mone03} indicates that
this separator is approximately 85\% accurate, but as mentioned above, 
it appears only weakly correlated with the SDSS morphology for our candidates.
This difference may be due to the fact that we preferentially select candidates
significantly fainter than the average magnitude of the USNO objects.
A small number of candidates ($\sim$30) have no identifiable counterpart in
USNO-B, and so have no available proper motion.  We have flagged 
these candidates by listing `$-99$' in place of their proper motions.  

\section{Follow-Up}\label{follow}
Following our candidate identification, we selected a group of targets 
without SDSS spectroscopy from DR4 that were visible from
the MDM Observatory in Tucson, Arizona during early October.  Using the 
CCDS spectrometer on the 2.4m Hiltner Telescope, we obtained low
signal-to-noise spectra, sufficient for identification purposes, 
for seven candidates.  We found three of these to be QSOs.
Only one of the objects for which we obtained spectra
falls into the SDSS area (0825-19933778), and while it was targeted as a QSO 
candidate by SDSS, our spectrum indicates it is probably a galaxy.  

From the cumulative
calculated probabilities, discussed in \S\ref{probAss} above, we would
expect $4.3$ of our targets to be quasars, so our finding three
genuine QSOs is consistent with our calculated probabilities.  A
list of our spectroscopic targets and their identifications 
is given in Table \ref{tblSpecResult}.  
None of the three objects that we have identified as QSOs 
appear in the most recent Veron catalog \citep{vero06}.

In addition to taking spectra, we also examine 
SDSS information for
a number of the candidates we selected for spectroscopy, since more data 
have become available in DR5.  We find SDSS information 
available for $81$ of our potential targets, of which 
$19$ are identified as quasars
by SDSS.  An additional $23$ were selected by SDSS as possible 
quasars, but have no public spectra in DR5.  
A list of these candidates, and the 
associated identifications, is found in Table
\ref{tblSdssIds}.  By again summing our calculated probabilities, we predict 
that $40.2$ candidates should be QSOs, which is again consistent with the
number of targets and known QSOs.  In conjunction with the
results of our observations, this larger sample affirms that our
probability calculation is performing reasonably well.


\section{Conclusions}
We have developed tight photometric selection criteria, allowing us to identify
a large number of QSO candidates across $\sim$8000 deg$^{2}$ of the sky.
Since the GALEX AIS survey will cover the entire sky, 
it should be possible to apply our
criteria to new data as they become available, identifying numerous
additional QSO candidates, eventually covering the entire sky above
$|b|=25^{\circ}$.  Candidates identified as QSOs in follow-up studies
should fall in $0\leq{\it z}\leq2$ \citep{bian05}.

The contamination of our catalog by the presence of a significant number of 
galaxies, apparent from a cursory examination of Tables \ref{tblSpecResult} and
\ref{tblSdssIds}, should not come as a surprise given the purpose of the GALEX 
survey.  The presence of a large number of galaxies means that a
significant source of contamination can be eliminated 
using good-quality imaging, without even the need for precise photometry.  This
is advantageous, since many of the point-source contaminants appear to be white
dwarfs, many of which could be eliminated using proper motions.

While the contamination of our catalog by galaxies and foreground stars
is significant, and presents a challenge for follow-up observations,
our algorithm selects numerous strong candidates.
We have already identified some of these as QSOs.  It is also possible
that the contaminants (i.e., non-QSOs) in our sample may 
themselves be of considerable scientific interest.  While we have not 
done extensive
analysis, it appears that a combination of our selection criteria and
good proper motions could allow the identification of a number of 
nearby white dwarfs.

It should be possible to 
select good targets for further study based on the calculated probabilities
and other information available from our catalog, including the presence of 
matches to one or more additional
catalogs and proper motions from USNO-B.
The effort required to follow up any significant fraction of the candidates
in the catalog will be
considerable, but application of the available information
should allow reasonable observational efficiency.  

\acknowledgments
We are grateful to Samir Salim for sharing his insights about the GALEX 
data products, as well as to the Space Telescope Science
Institute and MAST for providing convenient access to those data. We also wish
to thank Gordon Richards for his suggestion for an additional 
comparison, following the posting of our original manuscript on astro-ph.
This paper benefited greatly from a very thoughtful report by an anonymous
referee.
We owe tremendous thanks to the GALEX Collaboration and the 
United States Naval Observatory, without whose work ours 
would not have been possible.  This publication makes
use of data products from the Faint Images of the Radio Sky at Twenty-cm (FIRST)
survey, the Sloan Digital Sky Survey, and the Veron quasar catalog.  
It also makes use of data products from 
the Two Micron All Sky Survey, which is a joint project of the University of 
Massachusetts and the Infrared Processing and Analysis Center, California 
Institute of Technology, funded by the National Aeronautics and Space 
Administration and the National Science Foundation.  This paper is partially 
based on observations obtained with XMM-Newton, an ESA science mission with
instruments and contributions directly funded by ESA member states and NASA, 
and we have made use of the VizieR catalog access tool, CDS, Strasbourg, France.
Our work was supported by grant AST 04-52758 from the 
National Science Foundation (NSF).

\clearpage
\begin{table}
\begin{tabular}{|c|l|}
\tableline
Boundary Number & Boundary Criterion\\
\tableline
0 & $FUV-NUV\geq37.314\left(NUV-R\right)-70.70372$\\
1 & $FUV-NUV\geq\left(NUV-R\right)-0.5$\\
2 & $NUV-R\geq-0.895$\\
3 & $FUV-NUV\geq0.5$\\
4 & $FUV-NUV\leq4.343$\\
5 & $B-R\geq-0.9$\\
6 & $B-R\leq0.5$\\
\tableline
\end{tabular}
\caption{Color criteria applied to merged GALEX-USNO sources to select
candidates.  We require that each boundary be individually satisfied,
eliminating as much background as possible while still retaining a reasonable
number of candidates.  There are additional 
criteria applied to select candidates, as discussed in \S\ref{photo}.
\label{tblBounds}}
\end{table}

\clearpage

\clearpage

\clearpage

\clearpage

\clearpage

\clearpage

\clearpage

\clearpage

\clearpage

\clearpage
\begin{deluxetable}{|ccrrrrrrrrrrrrrrlccrr|}
\tabletypesize{\scriptsize}
\tablewidth{0pt}
\rotate
\tablecaption{Catalog of QSO candidates.  The
Flags field describes any matches to other catalogs, and
is coded thus: V (Veron), M (2MASS), F (FIRST) and X (XMMSL1).  The
columns labeled $\mu_{\alpha}$ and $\mu_{\delta}$
list the proper motions in the right ascension and declination directions,
respectively, in units of mas ${\rm yr}^{-1}$.
We find that USNO photometry is accurate to $\sim$0.35 mag.
The $\Delta$ values are listed in arcseconds, and
$\Delta_{A} \equiv A_{galex}-A_{usno}$.
Equatorial coordinates are listed in J2000 equinox.  The SDSS flag encoding
is described in Table \ref{tblSdssFlag}.\label{tblCand}}
\tablehead{
\colhead{UsnoA2} & \colhead{GalexId} & \colhead{$\alpha_{galex}$} & \colhead{$\delta_{galex}$} & \colhead{$\Delta_{\alpha}$} & \colhead{$\Delta_{\delta}$} & \colhead{B} & \colhead{S/G$_{B}$} & \colhead{R} & \colhead{S/G$_{R}$} & \colhead{$FUV$} & \colhead{$\sigma_{fuv}$} & \colhead{$NUV$} & \colhead{$\sigma_{nuv}$} & \colhead{$\mu_{\alpha}$} & \colhead{$\mu_{\delta}$} & \colhead{var} & \colhead{$SDSS$} & \colhead{Flags} & \colhead{$P_{ext}$} & \colhead{$P_{qso,ps}$}}
\startdata
0675-00000126 & 2668921591511910656 & 000.00337 & -21.29793 & +00 & +00 & 18.5 & 4.5 & 18.0 & 7.5 & 20.16 & 0.22 & 18.77 & 0.09 & +000.0 & +000.0 & no & 0 &  & 0.0094 & 0.6901\\ 
\tableline
0750-00000123 & 2674128848516287376 & 000.00822 & -09.78075 & +01 & +01 & 18.0 & 8.0 & 18.2 & 2.5 & 21.51 & 0.36 & 19.90 & 0.13 & +000.0 & +000.0 & no & 3 &  & 0.2633 & 0.4545\\ 
\tableline
1125-00000322 & 2503018566919390734 & 000.01206 & +23.14655 & +00 & -01 & 18.9 & 7.5 & 19.4 & 9.0 & 23.15 & 0.40 & 20.25 & 0.04 & -016.0 & -018.0 & no & 0 &  & 0.0138 & 0.7117\\ 
\tableline
0975-00000292 & 2690067364777559570 & 000.01908 & +10.28768 & -01 & +00 & 19.4 & 9.0 & 19.2 & 3.0 & 21.98 & 0.46 & 20.10 & 0.16 & -038.0 & +016.0 & no & 0 &  & 0.2714 & 0.7159\\ 
\tableline
0450-00000669 & 2671243738594936409 & 000.02583 & -37.69187 & +00 & +00 & 17.0 & 8.0 & 17.1 & 2.5 & 20.78 & 0.25 & 18.97 & 0.07 & +000.0 & +000.0 & no & 0 & M & 0.1467 & 0.6356\\ 
\tableline
0750-00000422 & 2667267895958962537 & 000.02755 & -07.86469 & -01 & +00 & 18.9 & 5.5 & 19.0 & 9.5 & 21.18 & 0.37 & 20.20 & 0.17 & +000.0 & +000.0 & no & 0 &  & 0.3096 & 0.7389\\ 
\tableline
0750-00000431 & 2434901622555410610 & 000.02801 & -09.70021 & +00 & -01 & 19.4 & 2.5 & 19.2 & 8.0 & 22.62 & 0.24 & 20.04 & 0.10 & -018.0 & -004.0 & no & 5 &  & 0.0191 & 0.7159\\ 
\tableline
0750-00000422 & 2667267891663998650 & 000.02815 & -07.86431 & +00 & +01 & 18.9 & 5.5 & 19.0 & 9.5 & 21.09 & 0.30 & 20.16 & 0.17 & +000.0 & +000.0 & no & 0 &  & 0.4035 & 0.5959\\ 
\tableline
0825-00000465 & 2667127184240411608 & 000.02816 & -00.78156 & +00 & +00 & 18.3 & 8.0 & 18.5 & 5.0 & 20.36 & 0.23 & 19.33 & 0.09 & -004.0 & +004.0 & no & 3 &  & 0.1159 & 0.6970\\ 
\tableline
0300-00000816 & 2671630758097980909 & 000.04020 & -54.94137 & +03 & +00 & 18.2 & 1.0 & 18.0 & 9.0 & 20.83 & 0.32 & 19.47 & 0.10 & +044.0 & -100.0 & no & 0 &  & 0.2032 & 0.7165\\ 
\tableline
\enddata
\end{deluxetable}

\clearpage
\begin{table}
\begin{tabular}{|c|l|}
\tableline
SDSS Flag & Description\\
\tableline
0 & Candidate Not Matched to SDSS DR4\\
\tableline
1 & SDSS Spectroscopic QSO\\
\tableline
2 & SDSS Spectroscopic Object, non QSO\\
\tableline
3 & SDSS Point Source in SDSS QSO Selection Area\\
\tableline
4 & SDSS Point Source Outside Selection Area\\
\tableline
5 & SDSS Extended Source\\
\tableline
\end{tabular}
\caption{A summary of the allowed values of the SDSS field in the
catalog and the meaning associated with each value.\label{tblSdssFlag}}
\end{table}

\clearpage
\begin{deluxetable}{|ccccccccccc|}
\tabletypesize{\scriptsize}
\tablewidth{0pt}
\rotate
\tablecaption{Targets of limited spectroscopic follow-up and resulting
identifications.  Candidates are listed in order of decreasing
conditional probability, $P_{qso,ps}$.
\label{tblSpecResult}}
\tablehead{
\colhead{UsnoA2} & \colhead{GalexId} & \colhead{$\alpha_{galex}$} & \colhead{$\delta_{galex}$} & \colhead{B} & \colhead{R} & \colhead{FUV} & \colhead{NUV} & \colhead{$P_{ext}$} & \colhead{$P_{qso,ps}$} & \colhead{identification}}
\startdata
0900-00965406 & 2730529405564617292 & 063.531706 & +06.530249 & 16.8 & 17.4 & 20.0181 & 18.9442 & 0.2439 & 0.8747 & QSO\\
\tableline
0825-00917727 & 2692213620064913973 & 061.029129 & -04.323432 & 17.2 & 17.7 & 19.3757 & 18.2044 & 0.0000 & 0.6032 & white dwarf\\
\tableline
1350-07170209 & 2683171214963246695 & 114.759581 & +51.452008 & 17.0 & 17.1 & 19.2899 & 18.3432 & 0.1616 & 0.5998 & QSO\\
\tableline
0825-19933778 & 2417063145906373262 & 351.085003 & -00.106850 & 16.9 & 17.0 & 18.4359 & 17.8182 & 0.1152 & 0.5922 & galaxy\\
\tableline
0825-01094092 & 2732851565532546868 & 069.449286 & -00.558338 & 17.2 & 17.8 & 19.0123 & 18.2720 & 0.0449 & 0.5811 & QSO\\
\tableline
1200-00733002 & 2681904581863016067 & 026.357883 & +31.230357 & 14.3 & 14.4 & 16.2930 & 15.5469 & 0.0714 & 0.4184 & star\\
\tableline
\enddata
\end{deluxetable}

\clearpage
\begin{deluxetable}{|ccrrrrrrrrc|}
\tablecaption{Candidates for spectroscopy with additional photometric
or spectroscopic information from SDSS DR5.  Candidates marked `target'
were selected by SDSS as potential QSOs, but have no DR5 spectroscopy.  
Candidates are listed in order of decreasing conditional
probability, $P_{qso,ps}$ \label{tblSdssIds}}
\rotate
\tabletypesize{\scriptsize}
\tablewidth{0pt}
\tablehead{
\colhead{UsnoA2} & \colhead{GalexId} & \colhead{$\alpha_{galex}$} & \colhead{$\delta_{galex}$} & \colhead{B} & \colhead{R} & \colhead{FUV} & \colhead{NUV} & \colhead{$P_{ext}$} & \colhead{$P_{qso,ps}$} & \colhead{note}}
\startdata
1350-07258531 & 2419666789440952962 & 119.334475 & +46.140664 & 18.5 & 18.9 & 21.1635 & 19.6432 & 0.1361 & 0.8533 & galaxy\\
\tableline
1125-05419661 & 2734505226725757226 & 118.906600 & +28.797934 & 18.5 & 19.1 & 20.6899 & 19.2935 & 0.0000 & 0.8348 & target\\
\tableline
0750-21608063 & 2674128831336416631 & 359.907848 & -09.286531 & 17.6 & 17.8 & 21.5457 & 19.5283 & 0.0000 & 0.8277 & target\\
\tableline
1275-07063429 & 2419385314464241297 & 116.859656 & +38.071564 & 18.9 & 19.0 & 21.0233 & 20.0472 & 0.4028 & 0.8188 & target\\
\tableline
1350-07251808 & 2419842711301397137 & 118.974739 & +46.958306 & 18.5 & 19.2 & 21.8050 & 20.7278 & 0.5273 & 0.8019 & target\\
\tableline
1275-07063032 & 2683874910994960717 & 116.842331 & +42.259684 & 18.8 & 19.0 & 20.7258 & 20.0884 & 0.4387 & 0.7818 & galaxy\\
\tableline
0825-19928229 & 2417098330278467604 & 350.726370 & -00.306123 & 18.5 & 18.8 & 20.9769 & 20.0031 & 0.3180 & 0.7660 & target\\
\tableline
0825-19934300 & 2417133514650558073 & 351.118259 & -00.878914 & 18.9 & 19.3 & 20.7641 & 20.1369 & 0.1220 & 0.7653 & QSO\\
\tableline
0825-19939176 & 2417133514650558688 & 351.438406 & -00.719253 & 18.7 & 19.1 & 21.4841 & 20.2718 & 0.2730 & 0.7652 & target\\
\tableline
0825-00954157 & 2692213611474977784 & 063.038867 & -05.675050 & 18.3 & 18.9 & 21.5428 & 20.1445 & 0.2634 & 0.7568 & white dwarf\\
\tableline
\enddata
\end{deluxetable}

\end{document}